\documentclass[aps,prl,showpacs,twocolumn,superscriptaddress,citeautoscript]{revtex4}
\usepackage{graphics}
\usepackage{graphicx}
\usepackage{amsmath}
\usepackage{amsfonts}
\usepackage{epsfig}
\usepackage{comment,braket}
\usepackage{color}

\def\beq{\begin{equation}}
\def\eeq{\end{equation}}
\def\vk{{\bf k}}

\begin{document}

\title{Theoretical prediction 
and spectroscopic fingerprints 
of an orbital transition in CeCu$_2$Si$_2$}
\author{L.~V.~Pourovskii}
\affiliation{Centre de Physique Th\'eorique, CNRS, \'Ecole Polytechnique, 91128 Palaiseau, France}
\affiliation{Swedish e-science Research Centre (SeRC), Department of Physics, Chemistry and Biology (IFM), Link\"oping University, Link\"oping, Sweden}
\author{P.~Hansmann}
\affiliation{Centre de Physique Th\'eorique, CNRS, \'Ecole Polytechnique, 91128 Palaiseau, France}
\author{M.~Ferrero}
\affiliation{Centre de Physique Th\'eorique, CNRS, \'Ecole Polytechnique, 91128 Palaiseau, France}
\author{A.~Georges}
\affiliation{Centre de Physique Th\'eorique, CNRS, \'Ecole Polytechnique, 91128 Palaiseau, France}
\affiliation{Coll\`ege de France, 11 place Marcelin Berthelot, 75005 Paris, France}
\affiliation{DPMC, Universit\'e de Gen\`eve, 24 quai Ernest Ansermet, CH-1211 Gen\`eve, Suisse}
 
\begin{abstract}

We show that the heavy-fermion compound CeCu$_2$Si$_2$ undergoes a transition between two regimes dominated by different crystal-field states. At low pressure $P$ and low temperature $T$ the Ce 4$f$ electron resides in the atomic crystal-field ground state, while at high $P$ or $T$, the electron occupancy and spectral weight is transferred to an excited crystal-field level that hybridizes more strongly with itinerant states. These findings result from first-principles  
dynamical-mean-field-theory calculations. We predict experimental signatures of this orbital transition in X-ray spectroscopy. The corresponding fluctuations may be responsible for the second high-pressure superconducting dome observed in this and similar materials.

\end{abstract}

\maketitle

CeCu$_2$Si$_2$, the first discovered heavy-fermion superconductor \cite{Steglich1979}, still generates a lot of interest due to the peculiar shape of the superconducting (SC) region in its pressure~($P$)-temperature~($T$) phase diagram. Superconductivity in this compound is observed in a 
wide range of pressures from 0 to 7 GPa with the SC critical temperature T$_c$ featuring two maxima: 
$T_c\approx 0.6$~K at  $P_c$=0.45~GPa, and $T_c\approx 2$~K at $P_c^*\approx$4.5~GPa~\cite{Bellarbi1984,Thomas1993,Seyfarth_epl_2012,Seyfarth2012}.
 This double-dome shape of the SC region has also been observed in iso-electronic CeCu$_2$Ge$_2$\cite{Vargoz1998} and differs from the  SC phases in other Ce `122'-type compounds (CePd$_2$Si$_2$\cite{Grosche1996}, CeRh$_2$Si$_2$\cite{Movshovich1996}), which exhibit a single-dome SC phase in a much narrower range of pressures around an antiferromagnetic (AFM) quantum critical point. By substituting 10\% of Si with Ge one may completely separate the two SC domes in CeCu$_2$Si$_2$\cite{Yuan2003}, 
thus 
suggesting that the SC domain in pure CeCu$_2$Si$_2$ is actually a merge 
of two SC phases with different origins. 

The maximum of the low-pressure SC dome has been unambiguously related to an AFM quantum critical point located at $P_c$. 
Indeed, specific heat measurements 
under small applied pressures in an external magnetic field~\cite{Lengyel2011} reveal that small deviations from the nominal stoichiometry stabilize either the AFM or SC phases 
at zero pressure \cite{Steglich1996}. The SC transition is accompanied by a lowering of the magnetic exchange energy\cite{Stockert2011}.
It is widely accepted, based on these observations, that the low-pressure SC phase is due to spin-fluctuation mediated pairing, 
similar to the single-dome SC in CePd$_2$Si$_2$ and CeRh$_2$Si$_2$. 

In contrast, no consensual 
picture has emerged up to date for the pairing mechanism in the high-pressure SC phase. The AFM order is already suppressed at pressures significantly below $P_c^*$, ruling out spin-fluctuation driven SC. 
For $P\gtrsim P_c^*$, the effective mass estimated from the AC specific heat  is significantly  reduced~\cite{Holmes2007}. The normal-state resistivity around $P_c^*$ is described by 
$\rho=\rho_0+AT^n$, with a large enhancement of $\rho_0$ and a non-Fermi liquid exponent $n\approx 1$~\cite{Jaccard2005}.
Recent multiprobe transport measurements clearly revealed the proximity of a critical point close to $P_c^*$~\cite{Seyfarth_epl_2012,Seyfarth2012}. 
%
It has been proposed \cite{Miyake2007} that $P_c^*$ is associated with the critical end-point of a first-order valence transition (VT), and that the  
associated critical fluctuations may provide the pairing mechanism in the high-pressure SC phase~\cite{Holmes2007}. 
Such a VT, at which  the Ce-$4f$ orbital occupancy $n_f$ jumps discontinuously, has been obtained within a single-band periodic Anderson model (PAM) in which 
an additional repulsion between the conduction electron band and the $f$-orbital is introduced~\cite{Onishi2000}. 
%
However, recent X-ray absorption measurements in a wide pressure range from $0$ to $7.8$~GPa detected only a smooth and weak decrease of $n_f$ 
as a function of pressure, without any marked feature around $P_c^*$~\cite{Rueff2011}. 
These results are in clear contradiction  
with the proposed valence transition and valence-fluctuation mechanism for SC.

\begin{figure}
\begin{center}
\includegraphics[width=0.95\columnwidth]{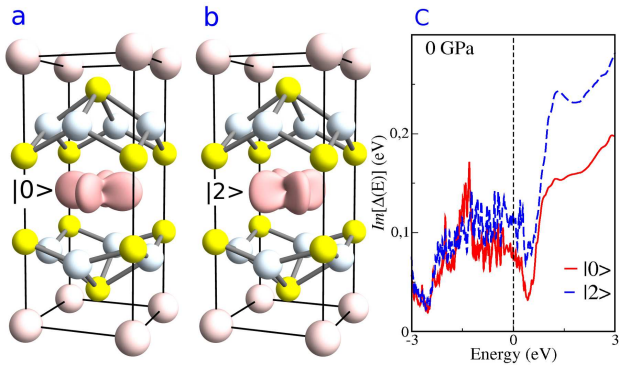}
\end{center}
\caption{\label{fig:struct_delta} (Color online)
(a,b). The CeCu$_2$Si$_2$ crystal structure. The pink, white, and yellow spheres are Ce, Cu, and  Si sites, respectively. 
(b) At the central Ce site, wavefunctions of the two CF levels are shown ($|0\rangle$ in (a) and $|2\rangle$ in (b)). 
(c) Imaginary part of the DMFT hybridization functions $\Delta$ of states $|0\rangle$ (red solid line) and $|2\rangle$ (blue dashed line) on the real energy axis at $P=0$~GPa.}
\end{figure}

In this letter, we provide theoretical evidence that $P_c^*$ is actually associated with an orbital transition between two different crystal-field levels. 
This conclusion is reached by performing first-principles calculations of CeCu$_2$Si$_2$ which combine electronic structure methods 
(density functional theory in the local density approximation - LDA) with a many-body treatment of the strong correlations in the 
Ce-$4f$ shell (dynamical mean-field theory - DMFT). 
%
%
We investigated the evolution of the electronic structure of the normal paramagnetic state 
as a function of applied pressure and temperature in the range $0<P<8$~GPa, $7<T<58$~K. 
Our calculations reveal that while $n_f$ 
remains close to unity within the whole range, the occupancies of different crystal-field (CF) levels within the Ce 4$f^1$ multiplet change drastically as function of $P$ and/or $T$. 
At low pressure and temperature the 4$f$ electron is mostly located at the ground-state level of the atomic Hamiltonian, 
while with increasing $P$ (and $T$) the electron weight is transferred to an excited level, which hybridizes more strongly with itinerant bands. 
The transition as a function of pressure becomes more drastic at low temperature, 
hinting at a 
quantum critical point at $P \approx$2.5~GPa, in rather close proximity to the maximum of the second SC dome. 
We show that the low-energy electronic structure is affected by this orbital transition, with the main Kondo resonance changing its orbital character. 
Finally, we predict distinctive signatures of this orbital transition in non-resonant inelastic X-ray scattering (NIXS) experiments. 
%

\begin{figure}
\begin{center}
\includegraphics[width=0.95\columnwidth]{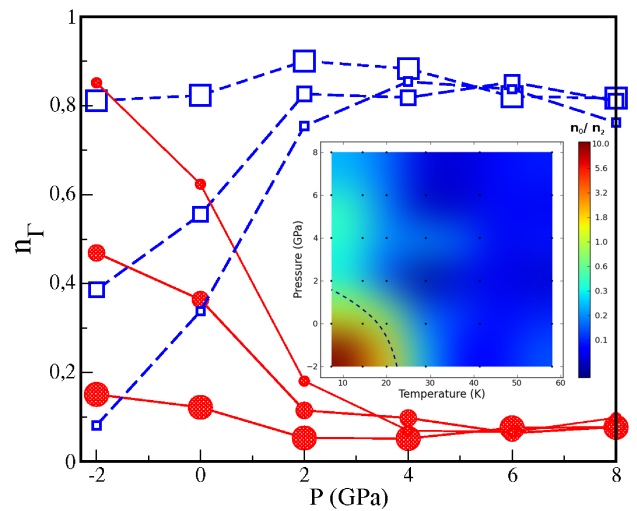}
\end{center}
\caption{\label{fig:n_Gamma} (Color online)
Occupancies $n_0$ (red circles) and $n_2$ (blue squares) of the CF states $|0\rangle$ and $|2\rangle$, as a function of pressure and temperature. 
The large, medium, and small symbols denote the occupancies at 58, 14, and 7~K, respectively. The curves are linear interpolations between the corresponding points. 
Inset: the $(T,P)$ map of the $n_0/n_2$ ratio. The dots indicate the values of $T$ and $P$ for which the LDA+DMFT calculations were performed. 
The dashed line is the $n_0=n_2$ boundary between the two regions, see text.}
\end{figure}

We use a fully charge self-consistent LDA+DMFT method~\cite{Aichhorn2009,Aichhorn2011} which combines a full-potential band-structure technique~\cite{Wien2k} 
with the DMFT~\cite{Georges1996} treatment of the on-site Coulomb repulsion between Ce 4$f$ states. 
The DMFT quantum impurity problem was solved with the numerically exact hybridization-expansion continuous-time quantum Monte-Carlo (CT-QMC) method~\cite{Gull2011}, 
as implemented in the TRIQS~\cite{TRIQS} package~\cite{METHOD_sup}. 


We calculated CeCu$_2$Si$_2$ in its experimental body-centered tetragonal ThCr$_2$Si$_2$-type structure (Fig.~\ref{fig:struct_delta}) and at the measured values of the lattice parameters vs.~$P$ reported in Ref.~\cite{Spain1986,note_negP}. In a tetragonal crystal field the $^2F_{5/2}$ ground-state multiplet of the Ce$^{3+}$ ion is split into three doublets: $|0\rangle=a|\pm 5/2\rangle+\sqrt{1-a^2}|\mp 3/2\rangle$, $|1\rangle=|\pm 1/2\rangle$, and $|2\rangle=\sqrt{1-a^2}|\pm 5/2\rangle-a|\mp 3/2\rangle$. As one sees in Fig~\ref{fig:struct_delta}{\it a} and {\it b}, the  CF states $|0\rangle$ and $|2\rangle$ differ by their orientation in the (001) plane: while the lobes of $|2\rangle$ point along [110] towards the nearest neighbor Si sites, the lobes of $|0\rangle$ point towards the neighboring Ce sites within the (001) plane. This difference in the spatial orientation leads to a stronger hybridization of $|2\rangle$ compared to $|0\rangle$, see Fig.~\ref{fig:struct_delta}c. 
When hybridization to the itinerant bands is neglected (e.g. in the Hubbard-I approximation), the splitting of CF levels 
exhibits a rather weak pressure dependence with  $|0\rangle$ being the ground state,  $|2\rangle$ the highest excited doublet and the total width of about 7 meV. 
This `bare' CF splitting is significantly smaller than the measured one of 30-37 meV\cite{Horn1981,Goremychkin1993,Ehm2007}, underlining the importance of hybridization effects 
in this compound.

\begin{figure*}
\begin{center}
\includegraphics[width=2.0\columnwidth]{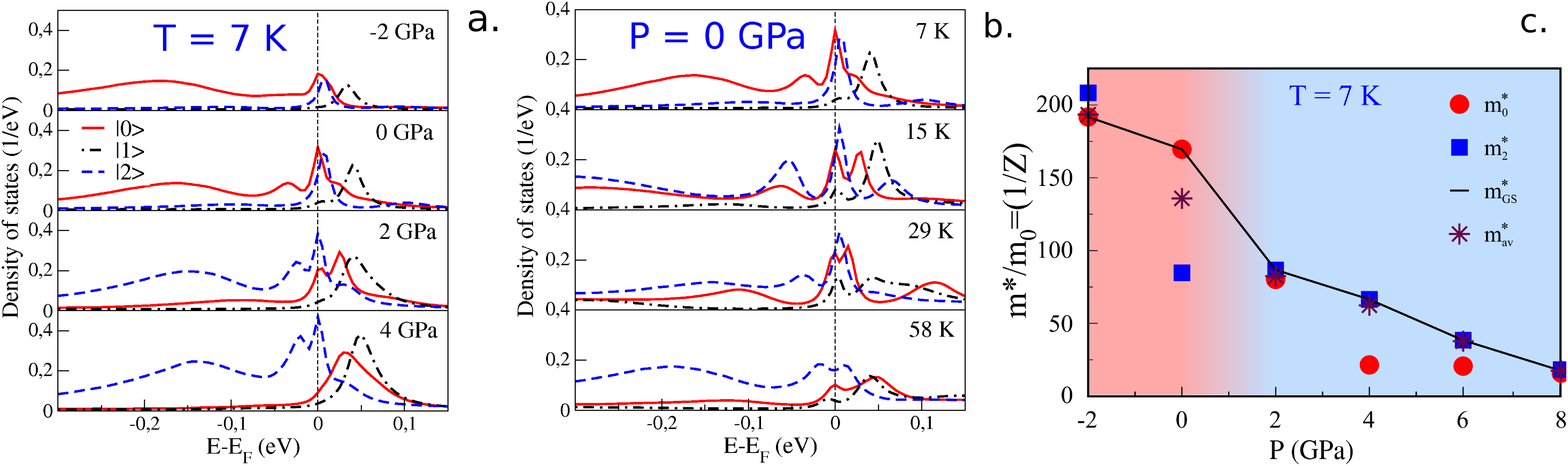}
\end{center}
\caption{\label{fig:DOS}
(Color online). (a,b) The LDA+DMFT partial densities of states of $|0\rangle$, $|1\rangle$ and  $|2\rangle$ states in the vicinity of the Fermi level as function of pressure at constant $T=$7~K (a) and as function of temperature at zero $P$ (b). The orbital character of the main Kondo peak changes as $P$ or $T$ increases. (c) The mass enhancements $m^*_{GS}$, $m_0$ and $m_2$ for the most-occupied, $|0\rangle$, and $|2\rangle$ states, respectively, as well as the average one  $m^*_{av}$ as function of $P$ at $T=$7~K. 
}
\end{figure*}

When the hybridization between Ce 4$f$ and itinerant electrons is fully included in the LDA+DMFT calculations using CT-QMC,  
the occupancies of the CF states $|0\rangle$ and $|2\rangle$ (designated by $n_0$ and $n_2$, respectively) develop a strong dependence on $P$ and $T$, which is displayed in Fig.~\ref{fig:n_Gamma} \cite{n1occ}. 
As shown there, at the highest $T=$58~K the strongly-hybridized state $|2\rangle$ dominates over the whole range of pressure. 
With lowering $T$ the occupancy $n_0$ increases for $P\lesssim$2~GPa at the expense of $n_2$, while at higher $P$ the occupancies exhibit almost no temperature dependence. 
As a result, at the lowest temperature $T=$7~K that we reached, the state $|0\rangle$ dominates at ambient and negative $P$ and its occupancy drops sharply between  0 and 2 GPa. 
In the inset of Fig.~\ref{fig:n_Gamma} we  map the ratio $n_0/n_2$ as a function of $P$ and $T$. The resulting `phase diagram' can be divided into two domains: the low $P$-low $T$  region with the  Ce 4$f$ mostly in the state $|0\rangle$ and the rest, where the state $|2\rangle$ dominates. The boundary $n_0=n_2$ between these two domains (dashed line in inset of Fig.~\ref{fig:n_Gamma}) extrapolated to $T=$0 gives $P_{cr}\approx$2.5~GPa. Recent {\bf q}-dependent NIXS measurements~\cite{Rueff2011} found Ce 4$f$ in the state $|2\rangle$ at ambient pressure and $T=$20~K, in agreement with our calculations. In contrast to the orbital occupancies, the total calculated occupancy of Ce 4$f$ shell shows modest dependence on pressure

At a qualitative level, this orbital transition can be captured by a periodic Anderson model consisting of two localized levels split by a CF field $\Delta_{cf}$, and such that 
the hybridization of the lowest level ($|0\rangle$) with itinerant bands is (approximately twice) smaller than that of the excited level ($|2\rangle$), as 
introduced in Ref.~\cite{Hattori2010}.  
The resulting orbital occupancy vs. ($V$,$T$) map~\cite{PAM_sup} for this model at low to moderate $T$ is remarkably similar to the one of CeCu$_2$Si$_2$ shown in Fig.~\ref{fig:n_Gamma}. 
We  note that the critical strength of hybridization $V_{cr}$ for the transition in the two-level PAM can be estimated from the condition 
$\Delta_{cf} = T_{\mathrm{K,ex}}-T_{\mathrm{K,gs}}\approx T_{\mathrm{K,ex}}$, where $T_{\mathrm{K,ex(gs)}}$ is the single-impurity Kondo scale for the excited (resp. ground-state) level and 
$T_{\mathrm{K,ex}} \gg T_{\mathrm{K,gs}}$ due to exponential dependence of $T_K$ on the hybridization strength.

The low-energy electronic structure of CeCu$_2$Si$_2$ is also affected by the orbital transition. In Fig.~\ref{fig:DOS}{\it a} we display the 
partial densities of states (PDOS, or orbital-resolved spectral functions) of the $|0\rangle$, $|1\rangle$ and  $|2\rangle$ orbitals in the vicinity of the Fermi level $E_F$ at $T=$7~K as a function of $P$. 
One sees that the Kondo peak due to the Ce 4$f$ quasiparticle states located at $E_F$ changes its orbital character from $|0\rangle$ to $|2\rangle$ as the system undergoes the orbital transition between 0 and 2 GPa. After the transition its weight rapidly increases with $P$ due to enhancement of the Kondo scale.  
The occupied spin-orbit and CF peaks at ambient $P$ are located at -0.2 eV and -35 meV, respectively, in agreement with recent photo-emission measurements~\cite{Ehm2007}. 
They are shifted to lower energies and change their orbital character as $P$ increases. The prominent CF satellite peaks above $E_F$ in Fig.~\ref{fig:DOS}{\it a} are due to empty (or weakly occupied) CF states, their positions with respect to Fermi level define the renormalized CF splitting. 
The calculated zero-pressure CF splitting of about 40 meV (the experimental value is 30-37 meV\cite{Horn1981,Goremychkin1993,Ehm2007}) 
exhibits a moderate increase with $P$, with the orbital character of the second CF peak switching across the transition.

\begin{figure}
\begin{center}
\includegraphics[width=1.0\columnwidth]{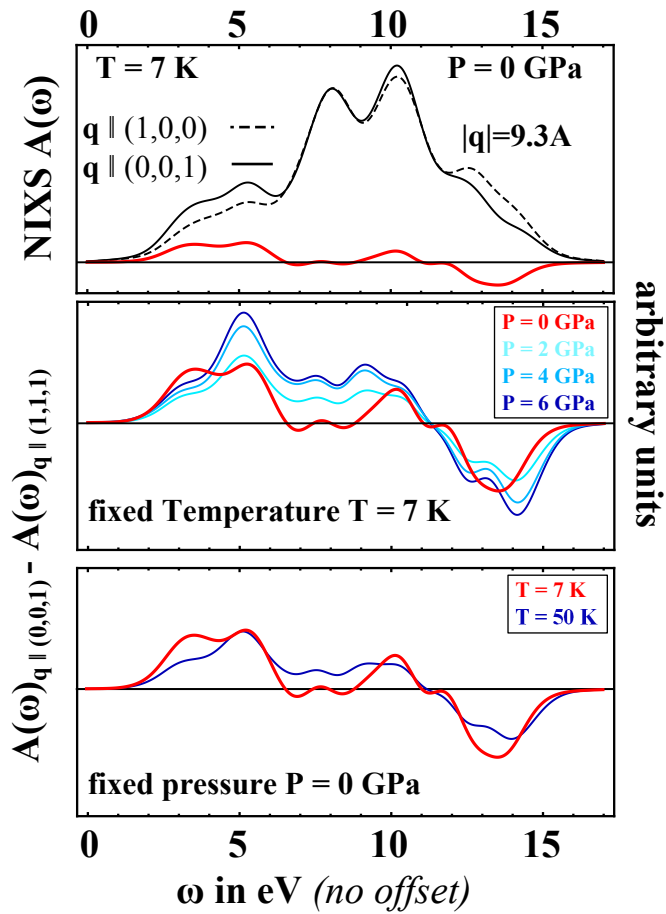}
\end{center}
\caption{\label{fig:exp}
(Color online). Simulation of NIXS crossections. In the top panel we show the spectral functions the momentum transfer dependent spectra (solid/dashed thin black line) and the difference sprectrum (thick red line). In the middle and bottom panels we display the NIXS difference spectrum upon increasing pressure at constant T=7~K and increasing temperature at ambient pressure, respectively. One may notice that the NIXS difference sprectrum undergoes structural changes due to the predicted orbital transition.
}
\end{figure}

The low-energy PDOS at ambient pressure and high $T=$58~K features a hump in the vicinity of $E_F$ due to the state $|2\rangle$. One may also notice there a small dip located exactly at $E_F$ (Fig.~\ref{fig:DOS}{\it b}), showing that a heavy-fermion Fermi-liquid state is not yet formed at this temperature. With decreasing $T$ the system moves to the phase where $|0\rangle$ dominates (inset of Fig.~\ref{fig:n_Gamma}), 
which is signaled by the formation of a Kondo peak of the corresponding character at $E_F$ and by a shift of the spectral weight of $|2\rangle$ away from $E_F$ to higher energies. The main Kondo peak appears to be fully formed only at the lowest $T$ of 7~K, in agreement with experimental estimates $T_K\approx$10~K for CeCu$_2$Si$_2$ at ambient $P$ \cite{Horn1981}.

We extracted the orbital-resolved mass enhancements $m^*_{\Gamma}$ from the corresponding self-energies $\Sigma_{\Gamma\Gamma}(i\omega)$ on the Matsubara grid as $1-\left[\left.d\mathrm{Im}\Sigma_{\Gamma\Gamma}(i\omega)/d\omega\right|_{\omega\to 0}\right]$, then the average mass enhancement $m^*_{av}$ was computed as $\sum_{\Gamma} m^*_{\Gamma}N_{\Gamma}(E_F)/\sum_{\Gamma}N_{\Gamma}(E_F)$, where $N_{\Gamma}(E_F)$ is the cor-
responding PDOS at the Fermi level. 
As shown on  Fig.~\ref{fig:DOS}{\it c}, the orbital transition $|0\rangle \rightarrow |2\rangle$ is accompanied by a significant reduction of the mass enhancement 
$m^*_{\mathrm{GS}}$ of the most occupied orbital (with $|\mathrm{GS}\rangle=|0\rangle$ for $P \leq 0$~GPa and $=|2\rangle$ for $P \geq 2$~GPa), 
while $m^*_{\mathrm{GS}}$ exhibits a rather slow linear decay away from the transition region. 
The evolution of the average mass enhancement $m^*_{av}$ is generally the same as that of  $m^*_{GS}$, apart from $P=$0~GPa, where the system at $T=$7~K seems to be still in an intermediate state and is expected to move to the $|0\rangle$-dominated phase at lower $T$'s (see inset in Fig.~\ref{fig:n_Gamma}). This calculated mass enhancement vs. $P$ evolution is in good agreement with experiment\cite{Holmes2007}. 
Hence, one may conclude that the system in a `heavy' Fermi-liquid state at low $P$  transforms into a `lighter' Fermi liquid through the orbital transition, at which the dominating CF state changes. 

Finally, we discuss spectroscopic signatures of the orbital transition, providing a direct experimental test of our theoretical  predictions in future experiments. 
In the past years it has been established that linear dichroism at the cerium M-edge of XAS (dipole transitions from $3d$ core- to $4f$ valence states)\cite{Hansmann2008, Willers2012A} or, even richer in information, $\mathbf{q}$-dependence (momentum-transfer) in NIXS (dipole, octopole, and triakontadipole from $4d$ core- to $4f$ valence states) \cite{Haverkort2007, Willers2012B} 
directly reflect the symmetry of the local Ce wavefunction.

Hence, using full multiplet cluster calculations, we have simulated XAS and NIXS signals following from our theory~\footnote{For the simulations we used the code of Ref.\cite{Haverkort12} and parameters for the Slater integrals and spin-orbit coupling same as in \cite{Hansmann2008}. For the NIXS spectra the radial part of the crosssection was taken from \cite{Willers2012A} }. While the XAS spectra can only probe the ground state composition by means of absolute contributions of $|J_z=5/2>$ and $|J_z=3/2>$ respectively, NIXS is capable of probing also their relative sign, i.e. the orientation of the wavefunction in the $ab$-plane, and hence distinguish states $|0\rangle$ and $|2\rangle$.
We thus refer to the supplementary material for the XAS spectra and focus here on the more informative momentum-transfer dependent NIXS signal which we report in Fig.\ref{fig:exp}.
The upper panel displays, at $P=0$ and $T=7$K, the spectral function for two different directions of momentum transfer (solid: $\mathbf{q}|| (0,0,1)$, dashed: $\mathbf{q}|| (1,0,0)$) at a fixed absolute value of $|\mathbf{q}|=9.3\text{\AA}^{-1}$. Also plotted in red/dark grey is the difference spectrum $A(\omega)_{\mathbf{q}|| (0,0,1)}-A(\omega)_{\mathbf{q}|| (1,0,0)}$.
The central panel displays the evolution of this difference spectrum at fixed temperature ($7$K) upon increasing pressure from $0$GPa to $6$GPa. While at $0$GPa the ground state is dominated by state $|0>$ (red spectrum) the switch to a state $|2>$ dominated ground state (light blue spectrum) already at $2$GPa is clearly visible, e.g. in the respective amplitude of the first two peaks between $3$ eV and $5$ eV, or in the increase between $7$ eV and $10$ eV.
We find the same clear-cut fingerprint of the orbital transition for the evolution at ambient pressure upon increasing temperature to $\approx 50$K (bottom panel). Also here the change of the ground state wavefunction is signaled by a change of the difference spectrum. While this change is qualitatively similar to the evolution with pressure, the absolute spectra differ due to the (slightly) different absolute values of the $|J_z=5/2>$ and $|J_z=3/2>$ coefficients.

In conclusion, our first-principles LDA+DMFT calculations predict the existence of a pressure- and temperature- induced orbital transition in CeCu$_2$Si$_2$. 
At this transition, the 4$f$ electron weight is transferred from CF state $|0\rangle$ (the atomic ground-state for vanishing hybridization) to the excited level $|2\rangle$, 
because the latter hybridizes more strongly with conduction electrons. Our results lead to clear-cut predictions for spectroscopic experiments like NIXS, 
where the fingerprint of the orbital transition can be detected in momentum-transfer dependent scattering cross sections. 
A similar "meta-orbital" transition between two CF levels with different hybridizations has been recently discussed in the context of a two-band periodical Anderson model~\cite{Hattori2010}. 
However, it has not been, to our knowledge, demonstrated from {\it ab initio} simulations of any real heavy-fermion material. 
It is tempting to speculate that the critical fluctuations associated with this orbital transition are responsible for the pairing in the high-pressure SC dome of 
CeCu$_2$Si$_2$ and iso-electronic CeCu$_2$Ge$_2$. Indeed, the calculated critical pressure of 2.5 GPa at zero temperature is rather close to the experimental maximum of this dome. 
Similar orbital transitions between CF levels may also explain the superconductivity away from magnetic quantum points in other HF compounds. 
It will be interesting to investigate whether such orbital transitions are always related to a double-dome SC or whether they can as well occur in single-dome and non-SC HF compounds (e.g. CePd$_2$Si$_2$,  CeAl$_2$).
We finally note that `composite pairing' SC has been proposed theoretically~\cite{Flint2008} to arise at the boundary between two distinct HF liquids originating in 
two orthogonal CF levels.  

We acknowledge discussions with D.~Jaccard, who attracted our attention to this problem, as well as with J.-P.~Rueff, T. Willers and M. Haverkort. 
Computing resources were provided by 
the Swedish National Infrastructure for Computing (SNIC) 
at the National Supercomputer Centre(NSC) and PDC Center for High Performance Computing, IDRIS-GENCI, 
and the  Swiss Center for Scientific Computing. 


\begin{thebibliography}{40}
\expandafter\ifx\csname natexlab\endcsname\relax\def\natexlab#1{#1}\fi
\expandafter\ifx\csname bibnamefont\endcsname\relax
  \def\bibnamefont#1{#1}\fi
\expandafter\ifx\csname bibfnamefont\endcsname\relax
  \def\bibfnamefont#1{#1}\fi
\expandafter\ifx\csname citenamefont\endcsname\relax
  \def\citenamefont#1{#1}\fi
\expandafter\ifx\csname url\endcsname\relax
  \def\url#1{\texttt{#1}}\fi
\expandafter\ifx\csname urlprefix\endcsname\relax\def\urlprefix{URL }\fi
\providecommand{\bibinfo}[2]{#2}
\providecommand{\eprint}[2][]{\url{#2}}

\bibitem[{\citenamefont{Steglich et~al.}(1979)\citenamefont{Steglich, Aarts,
  Bredl, Lieke, Meschede, Franz, and Sch\"afer}}]{Steglich1979}
\bibinfo{author}{\bibfnamefont{F.}~\bibnamefont{Steglich}},
  \bibinfo{author}{\bibfnamefont{J.}~\bibnamefont{Aarts}},
  \bibinfo{author}{\bibfnamefont{C.~D.} \bibnamefont{Bredl}},
  \bibinfo{author}{\bibfnamefont{W.}~\bibnamefont{Lieke}},
  \bibinfo{author}{\bibfnamefont{D.}~\bibnamefont{Meschede}},
  \bibinfo{author}{\bibfnamefont{W.}~\bibnamefont{Franz}}, \bibnamefont{and}
  \bibinfo{author}{\bibfnamefont{H.}~\bibnamefont{Sch\"afer}},
  \bibinfo{journal}{Phys. Rev. Lett.} \textbf{\bibinfo{volume}{43}},
  \bibinfo{pages}{1892} (\bibinfo{year}{1979}).

\bibitem[{\citenamefont{Bellarbi et~al.}(1984)\citenamefont{Bellarbi, Benoit,
  Jaccard, Mignot, and Braun}}]{Bellarbi1984}
\bibinfo{author}{\bibfnamefont{B.}~\bibnamefont{Bellarbi}},
  \bibinfo{author}{\bibfnamefont{A.}~\bibnamefont{Benoit}},
  \bibinfo{author}{\bibfnamefont{D.}~\bibnamefont{Jaccard}},
  \bibinfo{author}{\bibfnamefont{J.~M.} \bibnamefont{Mignot}},
  \bibnamefont{and} \bibinfo{author}{\bibfnamefont{H.~F.} \bibnamefont{Braun}},
  \bibinfo{journal}{Phys. Rev. B} \textbf{\bibinfo{volume}{30}},
  \bibinfo{pages}{1182} (\bibinfo{year}{1984}).

\bibitem[{\citenamefont{Thomas et~al.}(1993)\citenamefont{Thomas, Thomasson,
  Ayache, Geibel, and Steglich}}]{Thomas1993}
\bibinfo{author}{\bibfnamefont{F.}~\bibnamefont{Thomas}},
  \bibinfo{author}{\bibfnamefont{J.}~\bibnamefont{Thomasson}},
  \bibinfo{author}{\bibfnamefont{C.}~\bibnamefont{Ayache}},
  \bibinfo{author}{\bibfnamefont{C.}~\bibnamefont{Geibel}}, \bibnamefont{and}
  \bibinfo{author}{\bibfnamefont{F.}~\bibnamefont{Steglich}},
  \bibinfo{journal}{Physica B: Condensed Matter}
  \textbf{\bibinfo{volume}{186–188}}, \bibinfo{pages}{303 }
  (\bibinfo{year}{1993}), ISSN \bibinfo{issn}{0921-4526}.

\bibitem[{\citenamefont{Seyfarth
  et~al.}(2012{\natexlab{a}})\citenamefont{Seyfarth, R\"uetschi, Sengupta,
  Georges, and Jaccard}}]{Seyfarth_epl_2012}
\bibinfo{author}{\bibfnamefont{G.}~\bibnamefont{Seyfarth}},
  \bibinfo{author}{\bibfnamefont{A.-S.} \bibnamefont{R\"uetschi}},
  \bibinfo{author}{\bibfnamefont{K.}~\bibnamefont{Sengupta}},
  \bibinfo{author}{\bibfnamefont{A.}~\bibnamefont{Georges}}, \bibnamefont{and}
  \bibinfo{author}{\bibnamefont{Jaccard}}, \bibinfo{journal}{EPL}
  \textbf{\bibinfo{volume}{98}}, \bibinfo{pages}{17012}
  (\bibinfo{year}{2012}{\natexlab{a}}).

\bibitem[{\citenamefont{Seyfarth
  et~al.}(2012{\natexlab{b}})\citenamefont{Seyfarth, R\"uetschi, Sengupta,
  Georges, Jaccard, Watanabe, and Miyake}}]{Seyfarth2012}
\bibinfo{author}{\bibfnamefont{G.}~\bibnamefont{Seyfarth}},
  \bibinfo{author}{\bibfnamefont{A.-S.} \bibnamefont{R\"uetschi}},
  \bibinfo{author}{\bibfnamefont{K.}~\bibnamefont{Sengupta}},
  \bibinfo{author}{\bibfnamefont{A.}~\bibnamefont{Georges}},
  \bibinfo{author}{\bibfnamefont{D.}~\bibnamefont{Jaccard}},
  \bibinfo{author}{\bibfnamefont{S.}~\bibnamefont{Watanabe}}, \bibnamefont{and}
  \bibinfo{author}{\bibfnamefont{K.}~\bibnamefont{Miyake}},
  \bibinfo{journal}{Phys. Rev. B} \textbf{\bibinfo{volume}{85}},
  \bibinfo{pages}{205105} (\bibinfo{year}{2012}{\natexlab{b}}).

\bibitem[{\citenamefont{Vargoz and Jaccard}(1998)}]{Vargoz1998}
\bibinfo{author}{\bibfnamefont{E.}~\bibnamefont{Vargoz}} \bibnamefont{and}
  \bibinfo{author}{\bibfnamefont{D.}~\bibnamefont{Jaccard}},
  \bibinfo{journal}{Journal of Magnetism and Magnetic Materials}
  \textbf{\bibinfo{volume}{177–181, Part 1}}, \bibinfo{pages}{294 }
  (\bibinfo{year}{1998}).

\bibitem[{\citenamefont{Grosche et~al.}(1996)\citenamefont{Grosche, Julian,
  Mathur, and Lonzarich}}]{Grosche1996}
\bibinfo{author}{\bibfnamefont{F.}~\bibnamefont{Grosche}},
  \bibinfo{author}{\bibfnamefont{S.}~\bibnamefont{Julian}},
  \bibinfo{author}{\bibfnamefont{N.}~\bibnamefont{Mathur}}, \bibnamefont{and}
  \bibinfo{author}{\bibfnamefont{G.}~\bibnamefont{Lonzarich}},
  \bibinfo{journal}{Physica B: Condensed Matter}
  \textbf{\bibinfo{volume}{223–224}}, \bibinfo{pages}{50 }
  (\bibinfo{year}{1996}), ISSN \bibinfo{issn}{0921-4526}.

\bibitem[{\citenamefont{Movshovich et~al.}(1996)\citenamefont{Movshovich, Graf,
  Mandrus, Thompson, Smith, and Fisk}}]{Movshovich1996}
\bibinfo{author}{\bibfnamefont{R.}~\bibnamefont{Movshovich}},
  \bibinfo{author}{\bibfnamefont{T.}~\bibnamefont{Graf}},
  \bibinfo{author}{\bibfnamefont{D.}~\bibnamefont{Mandrus}},
  \bibinfo{author}{\bibfnamefont{J.~D.} \bibnamefont{Thompson}},
  \bibinfo{author}{\bibfnamefont{J.~L.} \bibnamefont{Smith}}, \bibnamefont{and}
  \bibinfo{author}{\bibfnamefont{Z.}~\bibnamefont{Fisk}},
  \bibinfo{journal}{Phys. Rev. B} \textbf{\bibinfo{volume}{53}},
  \bibinfo{pages}{8241} (\bibinfo{year}{1996}).

\bibitem[{\citenamefont{Yuan et~al.}(2003)\citenamefont{Yuan, Grosche, Deppe,
  Geibel, Sparn, and Steglich}}]{Yuan2003}
\bibinfo{author}{\bibfnamefont{H.~Q.} \bibnamefont{Yuan}},
  \bibinfo{author}{\bibfnamefont{F.~M.} \bibnamefont{Grosche}},
  \bibinfo{author}{\bibfnamefont{M.}~\bibnamefont{Deppe}},
  \bibinfo{author}{\bibfnamefont{C.}~\bibnamefont{Geibel}},
  \bibinfo{author}{\bibfnamefont{G.}~\bibnamefont{Sparn}}, \bibnamefont{and}
  \bibinfo{author}{\bibfnamefont{F.}~\bibnamefont{Steglich}},
  \bibinfo{journal}{Science} \textbf{\bibinfo{volume}{302}},
  \bibinfo{pages}{2104} (\bibinfo{year}{2003}).

\bibitem[{\citenamefont{Lengyel et~al.}(2011)\citenamefont{Lengyel, Nicklas,
  Jeevan, Geibel, and Steglich}}]{Lengyel2011}
\bibinfo{author}{\bibfnamefont{E.}~\bibnamefont{Lengyel}},
  \bibinfo{author}{\bibfnamefont{M.}~\bibnamefont{Nicklas}},
  \bibinfo{author}{\bibfnamefont{H.~S.} \bibnamefont{Jeevan}},
  \bibinfo{author}{\bibfnamefont{C.}~\bibnamefont{Geibel}}, \bibnamefont{and}
  \bibinfo{author}{\bibfnamefont{F.}~\bibnamefont{Steglich}},
  \bibinfo{journal}{Phys. Rev. Lett.} \textbf{\bibinfo{volume}{107}},
  \bibinfo{pages}{057001} (\bibinfo{year}{2011}).

\bibitem[{\citenamefont{Steglich et~al.}(1996)\citenamefont{Steglich,
  Gegenwart, Geibel, Helfrich, Hellmann, Lang, Link, Modler, Sparn, Büttgen
  et~al.}}]{Steglich1996}
\bibinfo{author}{\bibfnamefont{F.}~\bibnamefont{Steglich}},
  \bibinfo{author}{\bibfnamefont{P.}~\bibnamefont{Gegenwart}},
  \bibinfo{author}{\bibfnamefont{C.}~\bibnamefont{Geibel}},
  \bibinfo{author}{\bibfnamefont{R.}~\bibnamefont{Helfrich}},
  \bibinfo{author}{\bibfnamefont{P.}~\bibnamefont{Hellmann}},
  \bibinfo{author}{\bibfnamefont{M.}~\bibnamefont{Lang}},
  \bibinfo{author}{\bibfnamefont{A.}~\bibnamefont{Link}},
  \bibinfo{author}{\bibfnamefont{R.}~\bibnamefont{Modler}},
  \bibinfo{author}{\bibfnamefont{G.}~\bibnamefont{Sparn}},
  \bibinfo{author}{\bibfnamefont{N.}~\bibnamefont{Büttgen}},
  \bibnamefont{et~al.}, \bibinfo{journal}{Physica B: Condensed Matter}
  \textbf{\bibinfo{volume}{223–224}}, \bibinfo{pages}{1 }
  (\bibinfo{year}{1996}).

\bibitem[{\citenamefont{Stockert et~al.}(2011)\citenamefont{Stockert, Arndt,
  Faulhaber, Geibel, Jeevan, Kirchner, Loewenhaupt, Schmalzl, Schmidt, Si
  et~al.}}]{Stockert2011}
\bibinfo{author}{\bibfnamefont{O.}~\bibnamefont{Stockert}},
  \bibinfo{author}{\bibfnamefont{J.}~\bibnamefont{Arndt}},
  \bibinfo{author}{\bibfnamefont{E.}~\bibnamefont{Faulhaber}},
  \bibinfo{author}{\bibfnamefont{C.}~\bibnamefont{Geibel}},
  \bibinfo{author}{\bibfnamefont{H.~S.} \bibnamefont{Jeevan}},
  \bibinfo{author}{\bibfnamefont{S.}~\bibnamefont{Kirchner}},
  \bibinfo{author}{\bibfnamefont{M.}~\bibnamefont{Loewenhaupt}},
  \bibinfo{author}{\bibfnamefont{K.}~\bibnamefont{Schmalzl}},
  \bibinfo{author}{\bibfnamefont{W.}~\bibnamefont{Schmidt}},
  \bibinfo{author}{\bibfnamefont{Q.}~\bibnamefont{Si}}, \bibnamefont{et~al.},
  \bibinfo{journal}{Nature Physics} \textbf{\bibinfo{volume}{7}},
  \bibinfo{pages}{119 } (\bibinfo{year}{2011}).

\bibitem[{\citenamefont{Holmes et~al.}(2007)\citenamefont{Holmes, Jaccard, and
  Miyake}}]{Holmes2007}
\bibinfo{author}{\bibfnamefont{A.~T.} \bibnamefont{Holmes}},
  \bibinfo{author}{\bibfnamefont{D.}~\bibnamefont{Jaccard}}, \bibnamefont{and}
  \bibinfo{author}{\bibfnamefont{K.}~\bibnamefont{Miyake}},
  \bibinfo{journal}{Journal of the Physical Society of Japan}
  \textbf{\bibinfo{volume}{76}}, \bibinfo{pages}{051002}
  (\bibinfo{year}{2007}).

\bibitem[{\citenamefont{Jaccard and Holmes}(2005)}]{Jaccard2005}
\bibinfo{author}{\bibfnamefont{D.}~\bibnamefont{Jaccard}} \bibnamefont{and}
  \bibinfo{author}{\bibfnamefont{A.~T.} \bibnamefont{Holmes}},
  \bibinfo{journal}{Physica B: Condensed Matter}
  \textbf{\bibinfo{volume}{359–361}}, \bibinfo{pages}{333 }
  (\bibinfo{year}{2005}).

\bibitem[{\citenamefont{Miyake}(2007)}]{Miyake2007}
\bibinfo{author}{\bibfnamefont{K.}~\bibnamefont{Miyake}},
  \bibinfo{journal}{Journal of Physics: Condensed Matter}
  \textbf{\bibinfo{volume}{19}}, \bibinfo{pages}{125201}
  (\bibinfo{year}{2007}).

\bibitem[{\citenamefont{Onishi and Miyake}(2000)}]{Onishi2000}
\bibinfo{author}{\bibfnamefont{Y.}~\bibnamefont{Onishi}} \bibnamefont{and}
  \bibinfo{author}{\bibfnamefont{K.}~\bibnamefont{Miyake}},
  \bibinfo{journal}{Journal of the Physical Society of Japan}
  \textbf{\bibinfo{volume}{69}}, \bibinfo{pages}{3955} (\bibinfo{year}{2000}).

\bibitem[{\citenamefont{Rueff et~al.}(2011)\citenamefont{Rueff, Raymond,
  Taguchi, Sikora, Iti\'e, Baudelet, Braithwaite, Knebel, and
  Jaccard}}]{Rueff2011}
\bibinfo{author}{\bibfnamefont{J.-P.} \bibnamefont{Rueff}},
  \bibinfo{author}{\bibfnamefont{S.}~\bibnamefont{Raymond}},
  \bibinfo{author}{\bibfnamefont{M.}~\bibnamefont{Taguchi}},
  \bibinfo{author}{\bibfnamefont{M.}~\bibnamefont{Sikora}},
  \bibinfo{author}{\bibfnamefont{J.-P.} \bibnamefont{Iti\'e}},
  \bibinfo{author}{\bibfnamefont{F.}~\bibnamefont{Baudelet}},
  \bibinfo{author}{\bibfnamefont{D.}~\bibnamefont{Braithwaite}},
  \bibinfo{author}{\bibfnamefont{G.}~\bibnamefont{Knebel}}, \bibnamefont{and}
  \bibinfo{author}{\bibfnamefont{D.}~\bibnamefont{Jaccard}},
  \bibinfo{journal}{Phys. Rev. Lett.} \textbf{\bibinfo{volume}{106}},
  \bibinfo{pages}{186405} (\bibinfo{year}{2011}).

\bibitem[{\citenamefont{Aichhorn et~al.}(2009)\citenamefont{Aichhorn,
  Pourovskii, Vildosola, Ferrero, Parcollet, Miyake, Georges, and
  Biermann}}]{Aichhorn2009}
\bibinfo{author}{\bibfnamefont{M.}~\bibnamefont{Aichhorn}},
  \bibinfo{author}{\bibfnamefont{L.}~\bibnamefont{Pourovskii}},
  \bibinfo{author}{\bibfnamefont{V.}~\bibnamefont{Vildosola}},
  \bibinfo{author}{\bibfnamefont{M.}~\bibnamefont{Ferrero}},
  \bibinfo{author}{\bibfnamefont{O.}~\bibnamefont{Parcollet}},
  \bibinfo{author}{\bibfnamefont{T.}~\bibnamefont{Miyake}},
  \bibinfo{author}{\bibfnamefont{A.}~\bibnamefont{Georges}}, \bibnamefont{and}
  \bibinfo{author}{\bibfnamefont{S.}~\bibnamefont{Biermann}},
  \bibinfo{journal}{Phys. Rev. B} \textbf{\bibinfo{volume}{80}},
  \bibinfo{pages}{085101} (\bibinfo{year}{2009}).

\bibitem[{\citenamefont{Aichhorn et~al.}(2011)\citenamefont{Aichhorn,
  Pourovskii, and Georges}}]{Aichhorn2011}
\bibinfo{author}{\bibfnamefont{M.}~\bibnamefont{Aichhorn}},
  \bibinfo{author}{\bibfnamefont{L.}~\bibnamefont{Pourovskii}},
  \bibnamefont{and} \bibinfo{author}{\bibfnamefont{A.}~\bibnamefont{Georges}},
  \bibinfo{journal}{Phys. Rev. B} \textbf{\bibinfo{volume}{84}},
  \bibinfo{pages}{054529} (\bibinfo{year}{2011}).

\bibitem[{\citenamefont{Blaha et~al.}(2001)\citenamefont{Blaha, Schwarz,
  Madsen, Kvasnicka, and Luitz}}]{Wien2k}
\bibinfo{author}{\bibfnamefont{P.}~\bibnamefont{Blaha}},
  \bibinfo{author}{\bibfnamefont{K.}~\bibnamefont{Schwarz}},
  \bibinfo{author}{\bibfnamefont{G.}~\bibnamefont{Madsen}},
  \bibinfo{author}{\bibfnamefont{D.}~\bibnamefont{Kvasnicka}},
  \bibnamefont{and} \bibinfo{author}{\bibfnamefont{J.}~\bibnamefont{Luitz}},
  \emph{\bibinfo{title}{WIEN2k, An augmented Plane Wave + Local Orbitals
  Program for Calculating Crystal Properties}} (\bibinfo{publisher}{Techn.
  Universitat Wien, Austria, ISBN 3-9501031-1-2.}, \bibinfo{year}{2001}).

\bibitem[{\citenamefont{Georges et~al.}(1996)\citenamefont{Georges, Kotliar,
  Krauth, and Rozenberg}}]{Georges1996}
\bibinfo{author}{\bibfnamefont{A.}~\bibnamefont{Georges}},
  \bibinfo{author}{\bibfnamefont{G.}~\bibnamefont{Kotliar}},
  \bibinfo{author}{\bibfnamefont{W.}~\bibnamefont{Krauth}}, \bibnamefont{and}
  \bibinfo{author}{\bibfnamefont{M.~J.} \bibnamefont{Rozenberg}},
  \bibinfo{journal}{Rev. Mod. Phys.} \textbf{\bibinfo{volume}{68}},
  \bibinfo{pages}{13} (\bibinfo{year}{1996}).

\bibitem[{\citenamefont{Gull et~al.}(2011)\citenamefont{Gull, Millis,
  Lichtenstein, Rubtsov, Troyer, and Werner}}]{Gull2011}
\bibinfo{author}{\bibfnamefont{E.}~\bibnamefont{Gull}},
  \bibinfo{author}{\bibfnamefont{A.~J.} \bibnamefont{Millis}},
  \bibinfo{author}{\bibfnamefont{A.~I.} \bibnamefont{Lichtenstein}},
  \bibinfo{author}{\bibfnamefont{A.~N.} \bibnamefont{Rubtsov}},
  \bibinfo{author}{\bibfnamefont{M.}~\bibnamefont{Troyer}}, \bibnamefont{and}
  \bibinfo{author}{\bibfnamefont{P.}~\bibnamefont{Werner}},
  \bibinfo{journal}{Rev. Mod. Phys.} \textbf{\bibinfo{volume}{83}},
  \bibinfo{pages}{349} (\bibinfo{year}{2011}).

\bibitem[{\citenamefont{Ferrero and Parcollet}()}]{TRIQS}
\bibinfo{author}{\bibfnamefont{M.}~\bibnamefont{Ferrero}} \bibnamefont{and}
  \bibinfo{author}{\bibfnamefont{O.}~\bibnamefont{Parcollet}},
  \bibinfo{note}{"TRIQS: a Toolbox for Research on Interacting Quantum
  Systems", http://ipht.cea.fr/triqs}.

\bibitem[{MET()}]{METHOD_sup}
\bibinfo{note}{See Supplemental Material at for the calculational details}.

\bibitem[{\citenamefont{Spain et~al.}(1986)\citenamefont{Spain, Steglich,
  Rauchschwalbe, and Hochheimer}}]{Spain1986}
\bibinfo{author}{\bibfnamefont{I.}~\bibnamefont{Spain}},
  \bibinfo{author}{\bibfnamefont{F.}~\bibnamefont{Steglich}},
  \bibinfo{author}{\bibfnamefont{U.}~\bibnamefont{Rauchschwalbe}},
  \bibnamefont{and}
  \bibinfo{author}{\bibfnamefont{H.}~\bibnamefont{Hochheimer}},
  \bibinfo{journal}{Physica B+C} \textbf{\bibinfo{volume}{139–140}},
  \bibinfo{pages}{449 } (\bibinfo{year}{1986}), ISSN \bibinfo{issn}{0378-4363}.

\bibitem[{not()}]{note_negP}
\bibinfo{note}{The negative-pressure lattice parameters have been obtained by
  extrapolating data from Ref. [25].}

\bibitem[{\citenamefont{Horn et~al.}(1981)\citenamefont{Horn, Holland-Moritz,
  Loewenhaupt, Steglich, Scheuer, Benoit, and Flouquet}}]{Horn1981}
\bibinfo{author}{\bibfnamefont{S.}~\bibnamefont{Horn}},
  \bibinfo{author}{\bibfnamefont{E.}~\bibnamefont{Holland-Moritz}},
  \bibinfo{author}{\bibfnamefont{M.}~\bibnamefont{Loewenhaupt}},
  \bibinfo{author}{\bibfnamefont{F.}~\bibnamefont{Steglich}},
  \bibinfo{author}{\bibfnamefont{H.}~\bibnamefont{Scheuer}},
  \bibinfo{author}{\bibfnamefont{A.}~\bibnamefont{Benoit}}, \bibnamefont{and}
  \bibinfo{author}{\bibfnamefont{J.}~\bibnamefont{Flouquet}},
  \bibinfo{journal}{Phys. Rev. B} \textbf{\bibinfo{volume}{23}},
  \bibinfo{pages}{3171} (\bibinfo{year}{1981}).

\bibitem[{\citenamefont{Goremychkin and Osborn}(1993)}]{Goremychkin1993}
\bibinfo{author}{\bibfnamefont{E.~A.} \bibnamefont{Goremychkin}}
  \bibnamefont{and} \bibinfo{author}{\bibfnamefont{R.}~\bibnamefont{Osborn}},
  \bibinfo{journal}{Phys. Rev. B} \textbf{\bibinfo{volume}{47}},
  \bibinfo{pages}{14280} (\bibinfo{year}{1993}).

\bibitem[{\citenamefont{Ehm et~al.}(2007)\citenamefont{Ehm, H\"ufner, Reinert,
  Kroha, W\"olfle, Stockert, Geibel, and L\"ohneysen}}]{Ehm2007}
\bibinfo{author}{\bibfnamefont{D.}~\bibnamefont{Ehm}},
  \bibinfo{author}{\bibfnamefont{S.}~\bibnamefont{H\"ufner}},
  \bibinfo{author}{\bibfnamefont{F.}~\bibnamefont{Reinert}},
  \bibinfo{author}{\bibfnamefont{J.}~\bibnamefont{Kroha}},
  \bibinfo{author}{\bibfnamefont{P.}~\bibnamefont{W\"olfle}},
  \bibinfo{author}{\bibfnamefont{O.}~\bibnamefont{Stockert}},
  \bibinfo{author}{\bibfnamefont{C.}~\bibnamefont{Geibel}}, \bibnamefont{and}
  \bibinfo{author}{\bibfnamefont{H.~v.} \bibnamefont{L\"ohneysen}},
  \bibinfo{journal}{Phys. Rev. B} \textbf{\bibinfo{volume}{76}},
  \bibinfo{pages}{045117} (\bibinfo{year}{2007}).

\bibitem[{n1o()}]{n1occ}
\bibinfo{note}{The occupancy of the state $|1\rangle$ remains small
  ($n_1\approx$0.1 ) over the whole range of $P$ and $T$.}

\bibitem[{\citenamefont{Hattori}(2010)}]{Hattori2010}
\bibinfo{author}{\bibfnamefont{K.}~\bibnamefont{Hattori}},
  \bibinfo{journal}{Journal of the Physical Society of Japan}
  \textbf{\bibinfo{volume}{79}}, \bibinfo{pages}{114717}
  (\bibinfo{year}{2010}).

\bibitem[{PAM()}]{PAM_sup}
\bibinfo{note}{See Supplemental Material at for the periodic Anderson model
  calculations and the resulting orbital occupancy vs. ($V$,$T$) map}.

\bibitem[{\citenamefont{Hansmann et~al.}(2008)\citenamefont{Hansmann, Severing,
  Hu, Haverkort, Chang, Klein, Tanaka, Hsieh, Lin, Chen et~al.}}]{Hansmann2008}
\bibinfo{author}{\bibfnamefont{P.}~\bibnamefont{Hansmann}},
  \bibinfo{author}{\bibfnamefont{A.}~\bibnamefont{Severing}},
  \bibinfo{author}{\bibfnamefont{Z.}~\bibnamefont{Hu}},
  \bibinfo{author}{\bibfnamefont{M.~W.} \bibnamefont{Haverkort}},
  \bibinfo{author}{\bibfnamefont{C.~F.} \bibnamefont{Chang}},
  \bibinfo{author}{\bibfnamefont{S.}~\bibnamefont{Klein}},
  \bibinfo{author}{\bibfnamefont{A.}~\bibnamefont{Tanaka}},
  \bibinfo{author}{\bibfnamefont{H.~H.} \bibnamefont{Hsieh}},
  \bibinfo{author}{\bibfnamefont{H.-J.} \bibnamefont{Lin}},
  \bibinfo{author}{\bibfnamefont{C.~T.} \bibnamefont{Chen}},
  \bibnamefont{et~al.}, \bibinfo{journal}{Phys. Rev. Lett.}
  \textbf{\bibinfo{volume}{100}}, \bibinfo{pages}{066405}
  (\bibinfo{year}{2008}).

\bibitem[{\citenamefont{Willers
  et~al.}(2012{\natexlab{a}})\citenamefont{Willers, Adroja, Rainford, Hu,
  Hollmann, K\"orner, Chin, Schmitz, Hsieh, Lin et~al.}}]{Willers2012A}
\bibinfo{author}{\bibfnamefont{T.}~\bibnamefont{Willers}},
  \bibinfo{author}{\bibfnamefont{D.~T.} \bibnamefont{Adroja}},
  \bibinfo{author}{\bibfnamefont{B.~D.} \bibnamefont{Rainford}},
  \bibinfo{author}{\bibfnamefont{Z.}~\bibnamefont{Hu}},
  \bibinfo{author}{\bibfnamefont{N.}~\bibnamefont{Hollmann}},
  \bibinfo{author}{\bibfnamefont{P.~O.} \bibnamefont{K\"orner}},
  \bibinfo{author}{\bibfnamefont{Y.-Y.} \bibnamefont{Chin}},
  \bibinfo{author}{\bibfnamefont{D.}~\bibnamefont{Schmitz}},
  \bibinfo{author}{\bibfnamefont{H.~H.} \bibnamefont{Hsieh}},
  \bibinfo{author}{\bibfnamefont{H.-J.} \bibnamefont{Lin}},
  \bibnamefont{et~al.}, \bibinfo{journal}{Phys. Rev. B}
  \textbf{\bibinfo{volume}{85}}, \bibinfo{pages}{035117}
  (\bibinfo{year}{2012}{\natexlab{a}}).

\bibitem[{\citenamefont{Haverkort et~al.}(2007)\citenamefont{Haverkort, Tanaka,
  Tjeng, and Sawatzky}}]{Haverkort2007}
\bibinfo{author}{\bibfnamefont{M.~W.} \bibnamefont{Haverkort}},
  \bibinfo{author}{\bibfnamefont{A.}~\bibnamefont{Tanaka}},
  \bibinfo{author}{\bibfnamefont{L.~H.} \bibnamefont{Tjeng}}, \bibnamefont{and}
  \bibinfo{author}{\bibfnamefont{G.~A.} \bibnamefont{Sawatzky}},
  \bibinfo{journal}{Phys. Rev. Lett.} \textbf{\bibinfo{volume}{99}},
  \bibinfo{pages}{257401} (\bibinfo{year}{2007}).

\bibitem[{\citenamefont{Willers
  et~al.}(2012{\natexlab{b}})\citenamefont{Willers, Strigari, Hiraoka, Cai,
  Haverkort, Tsuei, Liao, Seiro, Geibel, Steglich et~al.}}]{Willers2012B}
\bibinfo{author}{\bibfnamefont{T.}~\bibnamefont{Willers}},
  \bibinfo{author}{\bibfnamefont{F.}~\bibnamefont{Strigari}},
  \bibinfo{author}{\bibfnamefont{N.}~\bibnamefont{Hiraoka}},
  \bibinfo{author}{\bibfnamefont{Y.~Q.} \bibnamefont{Cai}},
  \bibinfo{author}{\bibfnamefont{M.~W.} \bibnamefont{Haverkort}},
  \bibinfo{author}{\bibfnamefont{K.-D.} \bibnamefont{Tsuei}},
  \bibinfo{author}{\bibfnamefont{Y.~F.} \bibnamefont{Liao}},
  \bibinfo{author}{\bibfnamefont{S.}~\bibnamefont{Seiro}},
  \bibinfo{author}{\bibfnamefont{C.}~\bibnamefont{Geibel}},
  \bibinfo{author}{\bibfnamefont{F.}~\bibnamefont{Steglich}},
  \bibnamefont{et~al.}, \bibinfo{journal}{Phys. Rev. Lett.}
  \textbf{\bibinfo{volume}{109}}, \bibinfo{pages}{046401}
  (\bibinfo{year}{2012}{\natexlab{b}}).

\bibitem[{\citenamefont{Flint et~al.}(2008)\citenamefont{Flint, Dzero, and
  Coleman}}]{Flint2008}
\bibinfo{author}{\bibfnamefont{R.}~\bibnamefont{Flint}},
  \bibinfo{author}{\bibfnamefont{M.}~\bibnamefont{Dzero}}, \bibnamefont{and}
  \bibinfo{author}{\bibfnamefont{P.}~\bibnamefont{Coleman}},
  \bibinfo{journal}{Nature Physics} \textbf{\bibinfo{volume}{4}},
  \bibinfo{pages}{643 } (\bibinfo{year}{2008}).

\bibitem[{\citenamefont{Haverkort et~al.}(2012)\citenamefont{Haverkort,
  Zwierzycki, and Andersen}}]{Haverkort12}
\bibinfo{author}{\bibfnamefont{M.~W.} \bibnamefont{Haverkort}},
  \bibinfo{author}{\bibfnamefont{M.}~\bibnamefont{Zwierzycki}},
  \bibnamefont{and} \bibinfo{author}{\bibfnamefont{O.~K.}
  \bibnamefont{Andersen}}, \bibinfo{journal}{Phys. Rev. B}
  \textbf{\bibinfo{volume}{85}}, \bibinfo{pages}{165113}
  (\bibinfo{year}{2012}).

\bibitem[{\citenamefont{{Beach}}()}]{Beach2004}
\bibinfo{author}{\bibfnamefont{K.~S.~D.} \bibnamefont{{Beach}}},
  \bibinfo{note}{cond-mat/0403055}.

\bibitem[{\citenamefont{Hewson}(2003)}]{Hewson_book}
\bibinfo{author}{\bibfnamefont{A.~C.} \bibnamefont{Hewson}},
  \emph{\bibinfo{title}{The Kondo Problem to Heavy Fermions}}
  (\bibinfo{publisher}{Cambridge University Press, ISBN 0521599474},
  \bibinfo{year}{2003}).

\end{thebibliography}

\newpage
\mbox{}
\setcounter{figure}{0}
\setcounter{page}{1}
\setcounter{section}{0}
\renewcommand{\thepage}{S\arabic{page}}  
\renewcommand{\thesection}{S\arabic{section}}     
\renewcommand{\thefigure}{S\arabic{figure}}

{\bf\Large Supplementary Information}

\section{Method}

Our LDA+DMFT \cite{Aichhorn2009,Aichhorn2011} combines a full-potential linear augmented plain-wave electronic structure method \cite{Wien2k} with the dynamical
mean-field theory (DMFT) \cite{Georges1996} treatment of the on-site Coulomb repulsion between Ce 4$f$ states.
To solve the DMFT quantum impurity problem (QIP) we employed the numerically exact hybridization-expansion continuous-time quantum Monte-Carlo (CT-QMC) method \cite{Gull2011} as implemented in the TRIQS \cite{TRIQS} package. The Wannier orbitals representing Ce 4$f$ states were constructed from the Kohn-Sham states within the energy range from -12.4 to 5.4 eV 
using the projective approach of Ref.~\cite{Aichhorn2009}
The local Coulomb interaction between Ce 4$f$ electrons was approximated by a spherically-symmetric density-density form parametrized by the Slater parameter $F_0$=$U$=6.36~eV obtained by a constrain LDA method and Hund's rule coupling $J$=0.7~eV. The fully-localized-limit form of the double-counting correction term was employed throught.
We used the eigenstates $|\Gamma\rangle$ of Ce$^{3+}$ obtained by diagonalizing the {\it ab initio} crystal-field and spin-orbit Hamiltonian as a basis in CT-QMC. Rather small off-diagonal elements of the DMFT hybridization function in the $\{\Gamma\}$ basis were neglected. This approximation allowed us to treat the full Ce 4$f$ shell and to access low temperatures using the fast "segment picture" algorithm of CT-QMC \cite{Gull2011}. The DMFT QIP was solved using up to 3$\times$10$^{11}$ CT-QMC moves with a measurement performed after each 200 moves. The analytical continuation of imaginary-frequency self-energies to the real axis was carried out with a stochastic version of the Maximum Entropy method\cite{Beach2004}.

\section{Two-band periodic Anderson model}

In order to understand better the origin of the orbital transition in CeCu$_2$Si$_2$ we have solved a two-band periodic Anderson model (PAM) introduced in Ref.~\cite{Hattori2010} within DMFT and using the CT-QMC method to treat the quantum impurity problem. The model reads

\begin{equation}\label{PAM}
\begin{split}
H=&H_{cond}+\sum_{i\sigma\Gamma}V_{\Gamma}(c^{\dagger}_{i\sigma}f_{i\Gamma\sigma}+h.c.)+\sum_{i\sigma\Gamma}\varepsilon_{\Gamma}f^{\dagger}_{i\Gamma\sigma}f_{i\Gamma\sigma} \\
 &+U\sum_{i\Gamma}n_{i\Gamma\uparrow}n_{i\Gamma\downarrow}+U\sum_{i\sigma\sigma'}n_{i1\sigma}n_{i2\sigma'}, 
\end{split}
\end{equation}
where $c^{\dagger}_{i\sigma}$($c_{i\sigma}$) and  $f^{\dagger}_{i\Gamma\sigma}$($f_{i\Gamma\sigma})$ are the creation (annihilation) operators on the lattice site $i$ for conduction and localized states, respectively.  Two localized states $\Gamma=1,2$ on each lattice site have the bare level position $\varepsilon_{\Gamma}$ and hybridize with the strength  $V_{\Gamma}$ with the conduction band $H_{cond}=\sum_{\vk\sigma}\epsilon_{\vk}c^{\dagger}_{\vk\sigma}c_{\vk\sigma}$. The last two terms in (\ref{PAM}) are the intra and inter-orbital on-site Coulomb repulsions, respectively. In this model we neglect Hund's rule coupling, assuming it to be not important for the case of the localized band occupancy close to 1 and a small hybridization.   We consider the case when the exited crystal-field (CF) level $|2\rangle$ hybridizes stronger with the conduction band ($\varepsilon_{1} < \varepsilon_{2}$ and $V_{1} > V_{2}$) and, hence, hybridization (Kondo screening) effects compete with the CF splitting.

The corresponding local Green's functions (GF) (omitting the spin index) for the localilzed levels the Matsubara grid read

\begin{equation}\label{gf_alpha}
G_{\Gamma}(i\omega_n)=S_{\Gamma}+S^2_{\Gamma}V_{\Gamma}^{2}G_c(i\omega_n),
\end{equation}
where the conduction electrons' local GF is
\begin{equation}\label{gf_c}
G_{c}(i\omega_n)=\tilde{D}(i\omega_n+\mu-V_1^2S_{1}-V_2^2S_{2}).
\end{equation}
Here $\tilde{D}(E)=\int_{-\infty}^{\infty} \frac{D(\epsilon)}{E-\epsilon}d\epsilon$ is the Hilbert transform of the conduction-electron density of states, 
$S_{\Gamma}=(i\omega_n+\mu-\varepsilon_{\Gamma}-\Sigma_{\Gamma}(\omega_n))^{-1}$, $\Sigma_{\Gamma}(\omega_n)$ is the DMFT self-energy for level $\Gamma$, $\mu$ is the chemical potential fixing
the total occupancy of conduction and localized states to 2 electrons per site. 
As usual, the DMFT self-consitency condition is
$G_{\Gamma}=G^{imp}_{\Gamma}$, where the local GF of the lattice problem is given by (\ref{gf_alpha}) and the  $G^{imp}_{\Gamma}$ is obtained by solving  the corresponding DMFT quantum
impurity problem (in our case, with the CT-QMC method).

We chose a particle-hole symmetric semi-cilcular density of states of half-bandwidth $D=$1 for the conduction band as well as the position of the lowest local level $\varepsilon_{1}=$-2.5, 
the splitting between the levels $\Delta_{12}=\varepsilon_{2}-\varepsilon_{1}=$0.007 and $U=$6.0. As one may notice, the local-levels' parameters were chosen as to mimic the two competing CF states in CeCu$_2$Si$_2$.
We varied the hybridization strenght to simulate the effect of pressure, while keeping the ratio of the hybridization matrix elements $V_2/V_1=2$.

\subsection{Results}

\begin{figure}
\begin{center}
\includegraphics[width=1.0\columnwidth]{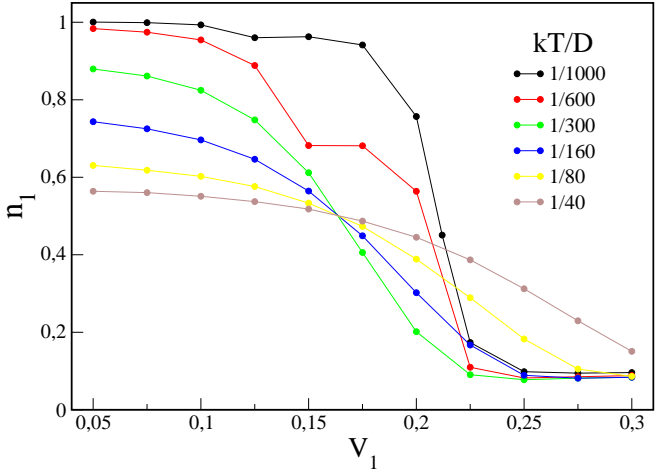}
\end{center}
\caption{\label{fig:n1_curves} The occupancy of the lowest level $n_1$ vs. the hybridization $V_1$ in the two-band PAM.
}
\end{figure}

In Fig.~\ref{fig:n1_curves} we display the evolution of the occupancy $n_1$ of the lowest level as function of the corresponding hybridization strenght $V_1$ for several values of the temperature $kT/D$. As one may notice, at high $T$ the occupancy $n_1$ decays slowly and smoothly with increasing $V_1$, while at low $T$ it exhibits a sharp drop from the values close to 1.0 to about 0.1 near the critical hybridization strenght $V_1^{cr}$=0.21 at $kT/D=$1000 (we obtained the value of $V_1^{cr}$ from the maximum of $|dn_1/dV_1|$). As in the case of CeCu$_2$Si$_2$, the total occupancy of the local levels $n_1+n_2$ features a rather weak decrease as function of $V$ even at low $T$ (it decreases by about 2\% from $V_1$=0.05 to $V_1$=0.30 at $\beta=$1000).

The critical value of hybridization can be obtained from a simple comparison of the energy gain due to the single-impurity Kondo screening with the value for the bare level splitting. We used for the single-impurity Kondo scale of the level $\Gamma$ the standard expression \cite{Hewson_book}
$$
T_{K\Gamma}=D(2J_{\Gamma}\rho)^{1/2}\exp\left(-\frac{1}{2J_{\Gamma}\rho}\right),
$$
where $J_{\Gamma}=\frac{V_{\Gamma}^2U}{|\varepsilon_{\Gamma}||\varepsilon_{\Gamma}+U|}$,  $D=$1 and $\rho=1$ are the half-bandwidth and the density of states at the Fermi level, respectively, for the  conduction states.  One may notice that for our range of parameters $T_{K2} \gg T_{K1}$. Then at the critical hybridization strenght $V_1^{cr}$ the gain in energy due to the Kondo screening $T_{K2}-T_{K1}\approx T_{K2}$ should be equal to the level splitting $\Delta_{12}$. Solving $T_{K2}=\Delta_{12}$ with our parameters gives $V_1^{cr}=0.207$, in very good agreement with the value obtianed in the full DMFT PAM calculations.

\begin{figure}
\begin{center}
\includegraphics[width=1.0\columnwidth]{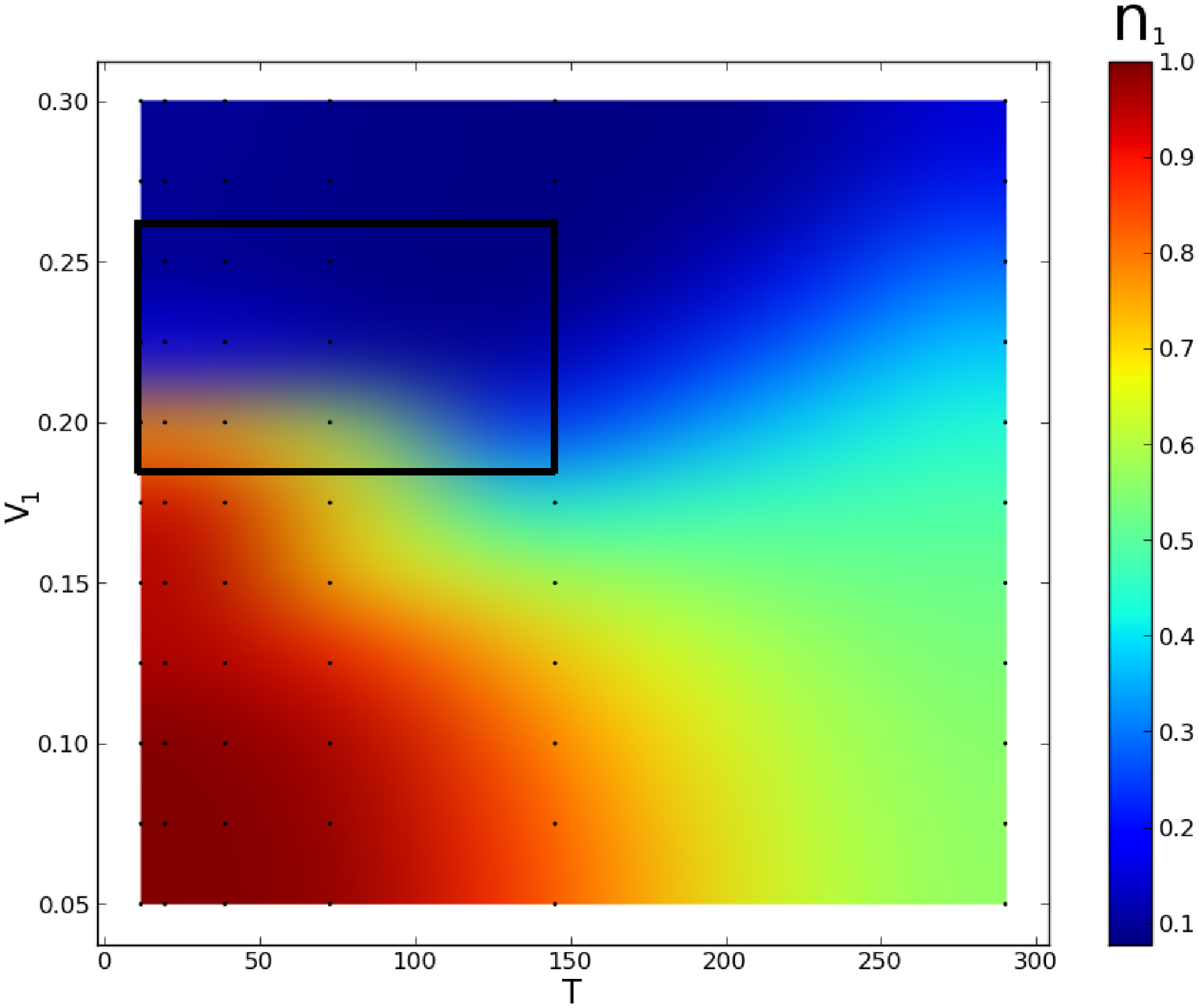}
\end{center}
\caption{\label{fig:n1_map} The  map occupancy of the lowest level $n_1$ as function of ($T$,$V_1$). The black dots indicate ($T$,$V_1$) for which DMFT calculations have been performed. The black rectangle in the upper left conner hihglights the region resembling the ($T$,$P$) map for CeCu$_2$Si$_2$ plotted in inset of Fig.~2 of the main text.
}
\end{figure}
In Fig.~\ref{fig:n1_map} we display the resulting map of the occupancy $n_1$ as function  of $T$ and $V_1$. To facilitate the comparison with the actual CeCu$_2$Si$_2$ case the temperature in Fig.~\ref{fig:n1_map} is plotted in Kelvins assuming the rest of model parameters in eV. As in CeCu$_2$Si$_2$ the low-temperature part can be divided in two regions with a rather sharp transition in between. In the first low-hybridization one the localized electron is at the lowest level $1$  and in the second large-$V$ region the electron is at the exited level $2$. The part of the diagram highlighted in Fig.~\ref{fig:n1_map} with the black rectangle is qualitatevely very similar to the corresponding ($T$,$P$) map for CeCu$_2$Si$_2$ plotted in inset of Fig.~2 of the main text, with a $n_1$ region located in the low-T/low-V coner and both hybridization and temperature-induced $n_1 \to n_2$ transitions possible. One may also notice that at high temperatures the orbital transition is smeared away.

\section{Calculations of X-ray spectra}

The importance of a reliable experimental determination of the symmetry of the CF ground state (by means of its $J_z$ composition) in heavy fermion compounds has been realized some time ago. First studies employed neutron scattering with great success in determining crystal field splittings with inelastic scattering. However, phonon scattering and magnetic excitations which occur in the same energy window prohibit a unambiguous determination of the exact ground state wave function.
Other attempts with fitting anisotropic magnetic susceptibility turned out to yield ambiguous results due to the sensitivity to non-local couplings. It turned out, however, that certain core level x-ray spectroscopy techniques, as a consequence of their intrinsic local nature, are able to yield the informations we seek (for further reading on this topic we refer to Refs.~\cite{Hansmann2008, Willers2012A,Willers2012B} and references therein). Hence, In order to test our hypothesis also experimentally, we have performed simulations of polarization dependent soft x-ray absorption spectroscopy (XAS) and momentum transfer dependent non-resonant x-ray scattering (NIXS).

The simulations have been performed by means of well established full Multiplet cluster calculations on a configuration interaction basis using the code of Ref.\cite{Haverkort12}. The most relevant excitations for both techniques which excite a (3d in XAS M-edge; 4d in NIXS) core electron have to be considered as very localized many body excitons. 

While linear dichroism is able to determine the $J_z$-composition by means of absolute values the sesitivity to the orientation of the orbital in the tetragonal $ab$-plane. It is only higher order transitions than dipole as measured in NIXS can distinguish between state $|0>$ and $|2>$. 
Below we give comprehensive ''maps'' of XAS and NIXS spectra as a function of temperature and pressure which should significantly simplify comparison of our results to experimental data.

\begin{figure*}
\begin{center}
\includegraphics[width=1.8\columnwidth]{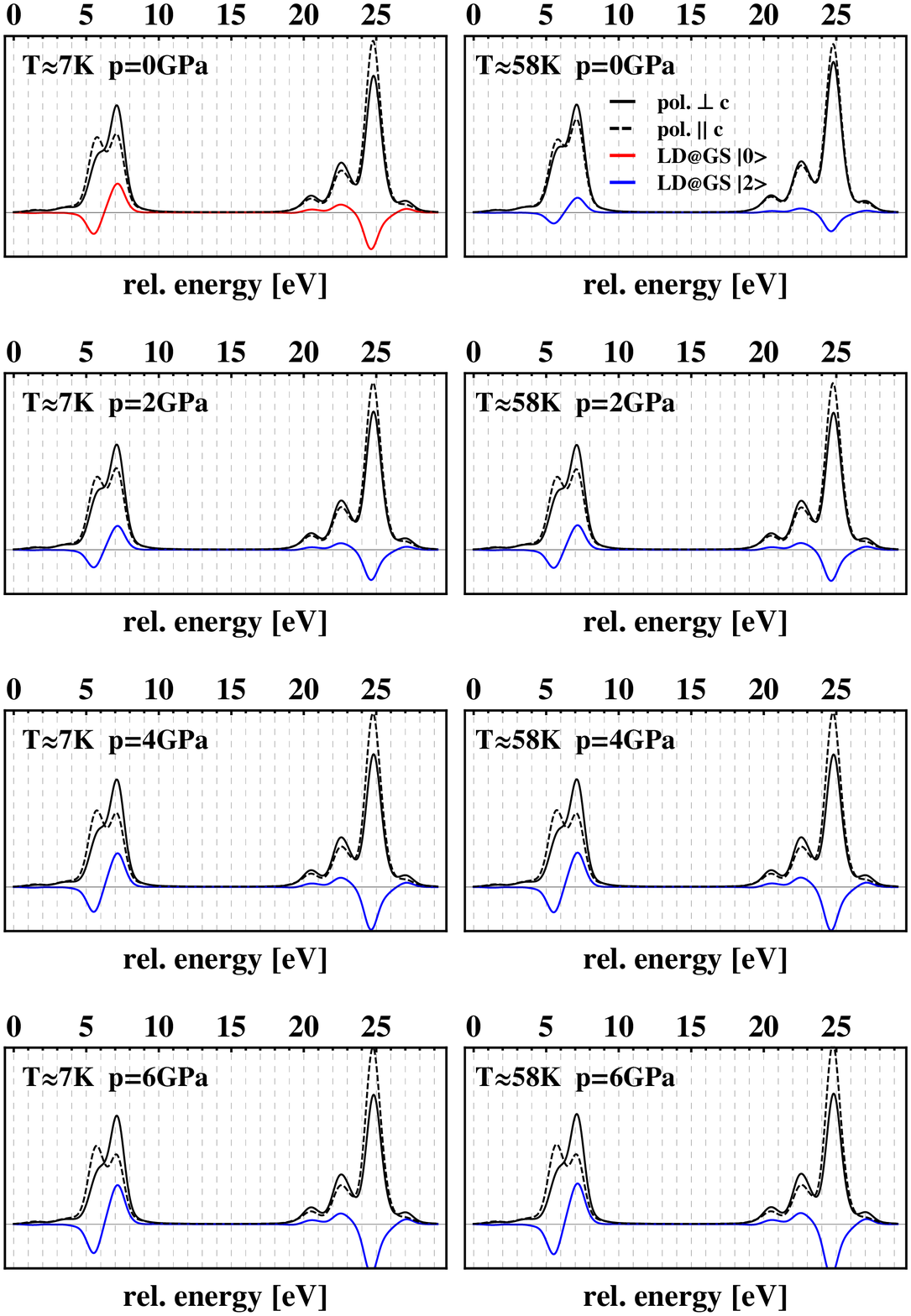}
\end{center}
\caption{\label{fig:expXAS}
(Color online). Simulation of x-ray absorption spectra and linear dichroism as a function of temperature and pressure. Plotted in black solid/dashed lines are the spectra for linear polarized light $\perp c$/$|| c$. The linear dichroism (LD) is plotted as red/blue curve for a $|0>$ / $|2>$ ground state.
}
\end{figure*}

\begin{figure*}
\begin{center}
\includegraphics[width=1.8\columnwidth]{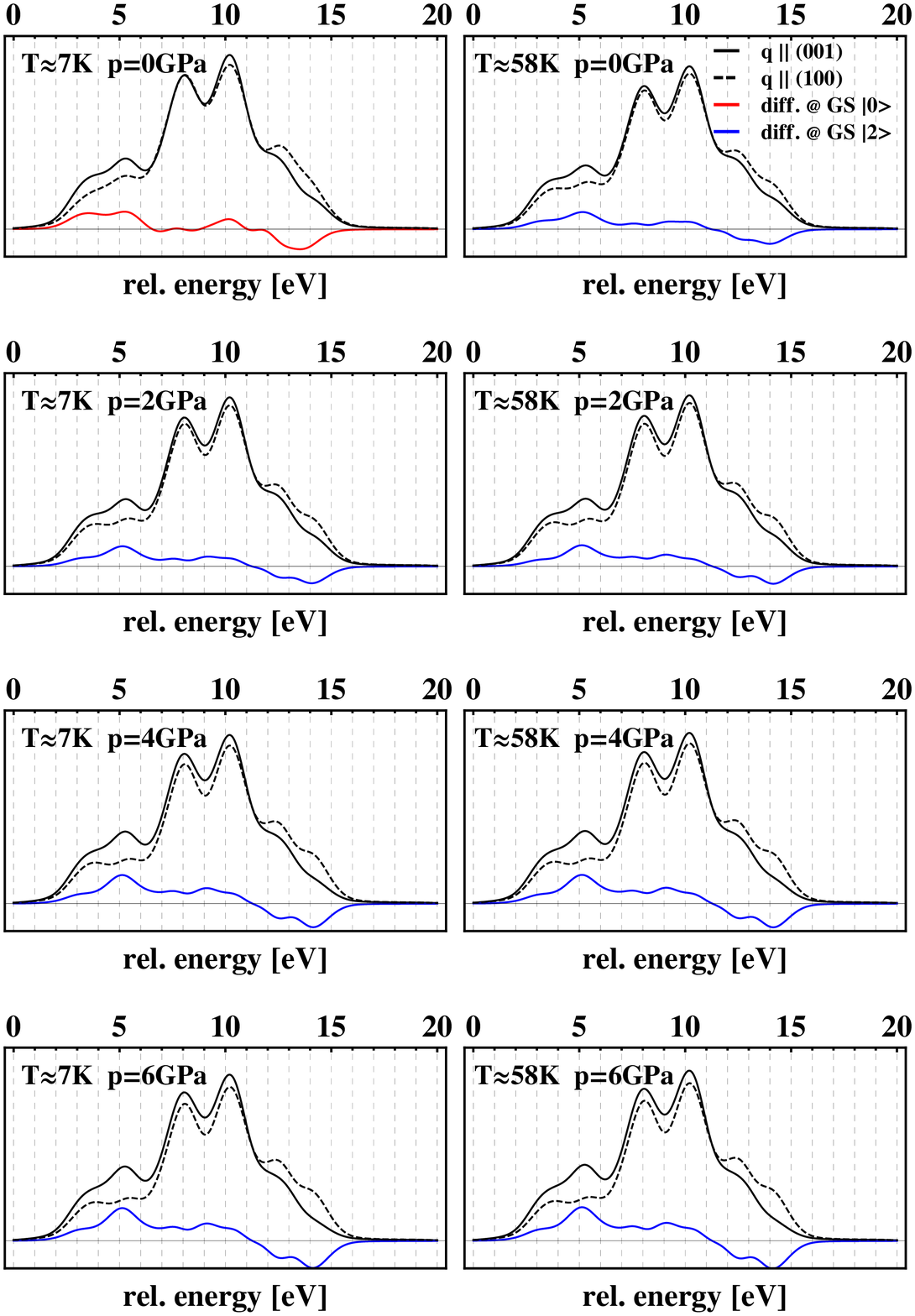}
\end{center}
\caption{\label{fig:expNIXS}
(Color online). Simulation of non resonant inelastic x-ray scattering (NIXS) and its momentum transfer dependence as a function of temperature and pressure. Plotted in black solid/dashed lines are the spectra for momentum transfer of $|\mathbf{q}|=9.3\AA$ in direction $\hat{\mathbf{q}} ||(001)$/$\hat{\mathbf{q}} ||(100)$. The difference spectrum is plotted as red/blue curve for a $|0>$ / $|2>$ ground state.
}
\end{figure*}

\end{document}